\documentclass[preprint]{elsarticle}
\usepackage{graphicx}
\usepackage{amssymb,amsmath}
\usepackage{subfig}
\usepackage{color}
\usepackage{xspace}
\newcommand{\nc}{\newcommand}
\nc{\rnc}{\renewcommand}
\nc{\x}[1]{\mbox{#1}}           
\nc{\hs}[1]{\hspace*{#1}}
\nc{\vs}[1]{\vspace*{#1}}
\rnc{\theta}{\vartheta}
\rnc{\phi}{\varphi}
\rnc{\rho}{\varrho}
\rnc{\epsilon}{\varepsilon}
\nc{\preload}{\Delta}
\nc{\scal}{\kappa}
\nc{\basefreq}{\Omega}
\nc{\red}[1]{\textcolor{red}{#1}}
\nc{\dd}{{\mathrm{d}}}
\nc{\herm}{^{\mathrm H}}
\nc{\ii}{{\mathrm{i}}}
\nc{\ee}{{\mathrm{e}}}
\nc{\mm}[1]{\mathbf{#1}}
\nc{\suml}[2]{\sum \limits_{#1}^{#2}}
\nc{\intl}[2]{\int \limits_{#1}^{#2}}
\rnc{\matrix}[2]{\left[\!\!\begin{array}{#1}
#2\end{array}\!\!\right]}
\rnc{\vector}[1]{\matrix{c}{#1}}
\nc{\inv}{^{-1}}
\nc{\tra}{^{\mathrm T}}
\nc{\sgn}{\mathrm{sgn}}
\nc{\kn}{{k_{\mathrm n}}}
\nc{\kt}{{k_{\mathrm t}}}
\nc{\xnl}{{\tilde x}}
\nc{\xc}{x_{\mathrm{c}}}
\nc{\fc}{F_{\mathrm R}}
\nc{\moptc}{m_{\mathrm{opt,c}}}
\nc{\ommod}{\omega_0}
\nc{\reliab}{RA}
\nc{\dmod}{D}
\nc{\amod}{q}
\nc{\stim}{\epsilon}
\nc{\areference}{a_{\mathrm{ref}}}
\nc{\prob}{\operatorname{Pr}}
\nc{\perf}{\Theta}
\nc{\phaslag}{\theta}
\nc{\perfref}{\Theta_{\mathrm{ref}}}
\nc{\fst}{F_0}
\nc{\nnn}{[1]}
\nc{\ttt}{[2]}
\nc{\ntd}{N_{\mathrm{td}}}
\nc{\kreis}[1]{\mbox{\mbox{\text{\Large{\ensuremath{\bigcirc}}}}\hspace{-2.4ex}#1\hspace{1.2ex}}}
\nc{\prog}[1]{{\sf{#1}}}
\nc{\matlab}{{Matlab}}
\nc{\name}[1]{{#1}}
\nc{\phaselagmethod}{{phase-resonance method}\xspace}
\nc{\petrovsmethod}{{horizontal-tangent method}\xspace}
\nc{\ie}{i.\,e.\,}
\nc{\eg}{e.\,g.\,}
\nc{\cf}{cf.\,}
\nc{\etal}{et~al.\,}
\nc{\z}[2]{\x{\sc #1}~{\x{\cite{#2}}}}
\nc{\zo}[1]{\x{\cite{#1}}}
\nc{\fabstand}{\,}
\nc{\fp}{\fabstand.}
\nc{\fk}{\fabstand,}
\nc{\real}[1]{\Re\lbrace #1 \rbrace}
\nc{\imag}[1]{\Im\lbrace #1 \rbrace}
\nc{\tab}[5][tbh]{\begin{table}[#1]\centering\caption{#4\label{tab:#5}}\begin{tabular}{#2}\hline #3 \\ \hline\end{tabular}\end{table}}
\newcommand{\fs}[4][tbh]{%
 \begin{figure}[#1]
 \centering
 \if@draft
  \framebox[150mm]{\raisebox{0mm}[25mm][25mm]{\texttt{#2}}}
 \else
  \includegraphics[scale=#4]{Figures/#2}
 \fi
 \caption{#3}
 \label{fig:#2}
\end{figure}
}
\newcommand{\fw}[3][tbh]{%
 \begin{figure}[#1]
 \centering
 \if@draft
  \framebox[150mm]{\raisebox{0mm}[25mm][25mm]{\texttt{#2}}}
 \else
  \includegraphics[width=1.0\textwidth]{Figures/#2}
 \fi
 \caption{#3}
 \label{fig:#2}
\end{figure}
}
\nc{\e}[2]{\begin{equation} #1 \label {eq:#2} \end{equation}}
\nc{\ea}[2]{
\begin{eqnarray}
#1 \label {eq:#2} \end{eqnarray}}
\nc{\eal}[3][0.0ex]{
\begin{samepage}
\begin{eqnarray*}
#2
\end{eqnarray*}
\nopagebreak[4] \vs{#1} \nopagebreak[4]\vs{-2ex} \nopagebreak[4]
\begin{eqnarray}
\label {eq:#3}
\end{eqnarray}
\end{samepage}\hs{-0.35em}}
\nc{\g}[1]{{$#1$}}
\nc{\fref}[1]{{Fig.~\ref{fig:#1}}}
\nc{\frefo}[1]{{\ref{fig:#1}}}
\nc{\frefoo}[1]{{#1}}
\nc{\frefs}[1]{{Figs.~\ref{fig:#1}}}
\nc{\tref}[1]{{Tab.~\ref{tab:#1}}}
\nc{\trefo}[1]{{\ref{tab:#1}}}
\nc{\trefs}[1]{{Tab.~\ref{tab:#1}}}
\nc{\erefn}[1]{{Eq.~(\ref{#1})}}
\nc{\eref}[1]{{Eq.~(\ref{eq:#1})}} \nc{\erefo}[1]{(\ref{eq:#1})}
\nc{\erefs}[1]{{Eqs.~(\ref{eq:#1})}}
\nc{\sref}[1]{{Section~\ref{sec:#1}}}
\nc{\srefo}[1]{\ref{sec:#1}}
\nc{\srefs}[1]{{Sections~\ref{sec:#1}}}
\nc{\aref}[1]{{{Appendix~\ref{asec:#1}}}}
\nc{\arefo}[1]{{\ref{asec:#1}}}
\nc{\arefs}[1]{{{Appendices~\ref{asec:#1}}}}
\nc{\ssref}[1]{{Subsection~\ref{sec:#1}}}
\nc{\ssrefo}[1]{\ref{sec:#1}}
\nc{\ssrefs}[1]{{Subsections~\ref{sec:#1}}}
\nc{\markit}[1]{#1}

\begin{document}

\begin{frontmatter}

\title{An efficient method for approximating resonance curves of weakly-damped nonlinear mechanical systems}

\author[ids]{Alwin F\"orster}
\author[ila]{Malte Krack\corref{cor1}}
\ead{malte.krack@ila.uni-stuttgart.de}

\address[ids]{Institute of Dynamics and Vibration Research,
Leibniz Universit\"at Hannover, 30167 Hannover, Germany}

\address[ila]{Institute of Aircraft Propulsion Systems,
University of Stuttgart, 70569 Stuttgart, Germany}

\cortext[cor1]{Corresponding author}

\begin{abstract}
\textit{A method is presented for tracing the locus of a specific peak in the frequency response under variation of a parameter. It is applicable to periodic, steady-state vibrations of harmonically forced nonlinear mechanical systems. It operates in the frequency domain and its central idea is to assume a constant phase lag between forcing and response. The method is validated for a two-degree-of-freedom oscillator with cubic spring and a bladed disk with shroud contact. The method provides superior computational efficiency, but is limited to weakly-damped systems. Finally, the capability to reveal isolated solution branches is highlighted.}
\end{abstract}


\end{frontmatter}


\section{Introduction\label{sec:introduction}}
The dynamical behavior of mechanical systems are governed by differential equations that are, in general, nonlinear in the describing coordinates. Once the nonlinear terms in the differential equations become relevant in the considered dynamical regime, we refer to these systems as nonlinear systems. Examples are systems with contact or dry friction, fluid-structure-interaction or large deflections. The nonlinear character makes their design and analysis more difficult and usually necessitates for appropriate numerical procedures. In the design process, it is often of interest to predict the vibration behavior in certain ranges of parameters. In the presence of sustained external forcing, the phenomenon of resonance is of particular concern. In this case, the steady-state vibrations can reach high levels, which may lead to structural damage and noise. To avoid resonances, or to ensure that the resonant vibration level is tolerable, it is important to predict resonance frequencies and associated vibration amplitudes. \markit{Throughout this work, only numerical methods are addressed, since the different analytical techniques are strictly limited in their range of applicability. Furthermore, we discuss only methods capable of analyzing the dependence on generic parameters. In contrast, methods limited to the analysis of only specific parameters are excluded from the discussion. An important example are nonlinear modal analysis methods \cite{kers2009}, which are only capable of revealing the dependence of resonances on the excitation level.}\\
The most straightforward way to obtain resonance frequencies and amplitudes is the computation of individual frequency responses for a discrete set of parameters. This technique often requires a large number of frequency response computations, because the system behavior can exhibit regimes of high or low sensitivities, and the associated parameter ranges are not a priori known. This might result in prohibitive computational effort.\\
An alternative technique was originally proposed by \name{Petrov} \zo{petr2006b}, and has been frequently applied since then, \eg in \zo{petr2009a}. The method aims at reducing the computational effort compared to crude forced response computations by directly determining the resonance curves, \ie, the curves that trace the locus of a specific peak of the frequency response under the variation of a parameter. To this end, a so-called resonance condition is introduced to the problem formulation, which requires the solution point to have a horizontal tangent in the amplitude-frequency plane. To this end, the unknown resonance frequency becomes part of the sought solution. The \textit{\petrovsmethod} is formulated in the frequency domain in the framework of the high-order harmonic balance method. The extended set of equations is solved using a gradient-based method in conjunction with a predictor-corrector arc-length continuation method. Since the computation of the tangent requires already the first-order derivatives, the gradient-based solution requires second-order derivatives of the residual with respect to the unknowns. This method can be categorized as a \textit{first-order method}, in the sense that the formulation of the problem already involves first-order derivatives. Computing the required second-order derivatives can result in a huge computational burden, in particular if they are approximated by a finite difference scheme. But even in the case of analytical derivatives, the plain evaluation of their algebraic forms involves a considerable overhead. If this overhead is in the order of magnitude of the potential computational savings achieved by the direct analysis of resonances, this overhead can render this technique useless. Moreover, the analytical calculation of second-order derivatives can become an exhaustive and tedious task in the case of generic nonlinearities. For the class of piecewise polynomial systems, fortunately, automated frameworks are available \zo{krac2013b}.\\
In order to gain further computational savings in the computation of resonance curves, we propose a \textit{zeroth-order method} in this work, \ie, a method that does not involve any derivatives in the formulation of the governing equations. The criterion is based on the phase lag between response and forcing. To this end, the notion of phase must be compatible with the analysis method, and we therefore place the method in the frequency domain framework.\\
In this paper we present this new method, which we refer to as \phaselagmethod. We recap the harmonic balance method and present the additional phase lag criterion in \ssref{hbm} and \ssref{phaselag}, respectively. In addition we show a qualitative analysis of the new method in comparison to the \petrovsmethod in \ssref{qualitative}. In \sref{application_twodof} and \sref{application_bladeddisk}, the method is validated for two nonlinear mechanical systems, namely a two degree-of-freedom (2DOF) oscillator with cubic spring, and a bladed disk with nonlinear shroud contact, respectively. Special attention is paid to the strongly nonlinear regime, where it is demonstrated that the method may also be useful to gain insight into the complicated behavior of dynamical systems.

\section{A novel method for the direct computation of forced resonances of nonlinear systems\label{sec:method}}
\nc{\myquote}[1]{`#1'}
\nc{\ndof}{N_{\mathrm{DOF}}}
\nc{\fex}{\mm{f}_{\mathrm{ex}}}
\nc{\fnl}{\mm{f}_{\mathrm{nl}}}
\nc{\nh}{N_{\mathrm h}}
    \subsection{Recap of the harmonic balance method\label{sec:hbm}}
    	The harmonic balance method\footnote{In the literature, other widely
used names for the method described here are the Describing Function
method and the method of Krylov-Bogoliubov-Mitropolsky. Moreover,
the prefix \myquote{multi} or \myquote{high-order} are often used
for the harmonic balance method in order to clarify the difference
to the single-term variant which only considers the fundamental
harmonic.} is a widely-used method to numerically compute the periodic, steady state oscillations of nonlinear dynamical systems. Consider a nonlinear mechanical system that has already been discretized in space and is described in terms of $\ndof$ generalized coordinates assembled in the vector $\mm{q}$. The motions $\mm{q}(t)$ of the system are governed by a set of ordinary differential equations,
    	\e{
    		 \mm{M}\ddot{\mm{q}}\left(t\right)+\mm{C}\dot{\mm{q}}\left(t\right)+\mm{K}\mm{q}\left(t\right)+\fnl\left(\mm{q}\left(t\right),\mm{\dot q}\left(t\right), \lambda\right)-\fex\left(t,\lambda\right)=\mm{0}\,,
    	}{EquationOfMotion_gen}
    	and appropriate initial conditions $\mm{q}(0) = \mm{q}_0$ and $\dot{\mm{q}}(0)={\mm u}_0$. In \eref{EquationOfMotion_gen} $\mm{M}$ is the mass matrix, $\mm{C}$ is the damping matrix, $\mm{K}$ is the stiffness matrix, $\fex \left(t,\lambda\right)$ is the vector of excitation forces, and $\fnl$ is the vector of nonlinear forces. Without loss of generality, it is assumed that $\fnl$ is essentially nonlinear in $\mm q$, $\mm{\dot q}$ so that the linearized behavior is fully accounted for in the structural matrices $\mm K$ and $\mm C$, respectively. Furthermore $t$ is the time and $\lambda$ is a parameter which may influence the excitation and/or the nonlinear force. It should be noted that the approach can also be applied if the structural matrices are parameter dependent. This case is merely excluded from the further development for the sake of an easier presentation.\\
    	In the harmonic balance method, periodic oscillations of the generalized coordinates $\mm{q}\left(t\right)$ with the fundamental frequency $\omega$ are assumed. The unknown coordinates are approximated by a truncated \name{Fourier} series with the harmonic order $\nh$:
    	\e{
    		\mm{q}\left(t\right)\approx\mm{Q}_{0}+\sum_{n=1}^{\nh}\mm{Q}_{n}^{\left(c\right)}\cos\left(n\omega t\right)+\mm{Q}_{n}^{\left(s\right)}\sin\left(n\omega t\right)\fp
    	}{Fourier_q}
    	The generalized velocities and accelerations are determined by deriving \eref{Fourier_q} with respect to time,
    	\ea{
    		\dot{\mm{q}}\left(t\right) &\approx& \sum_{n=1}^{\nh}-\mm{Q}_{n}^{\left(c\right)}n\omega\sin\left(n\omega t\right)+\mm{Q}_{n}^{\left(s\right)}n\omega\cos\left(n\omega t\right)\fk
    	\label{eq:Fourier_dq}\\
    		\ddot{\mm{q}}\left(t\right) &\approx& \sum_{n=1}^{\nh}-\mm{Q}_{n}^{\left(c\right)}n^{2}\omega^{2}\cos\left(n\omega t\right)-\mm{Q}_{n}^{\left(s\right)}n^{2}\omega^{2}\sin\left(n\omega t\right)\fp
    	}{Fourier_ddq}
    	For convenience, the Fourier coefficients are assembled in the following vectors:
    	\ea{
    		\tilde{\mm{Q}} &=&
    		\begin{pmatrix}
    			\left(\mm{Q}_{0}\right)\tra && \left(\mm{Q}_{1}^{(\mathrm{c})}\right)\tra && \left(\mm{Q}_{1}^{(\mathrm{s})}\right)\tra && \hdots && \left(\mm{Q}_{\nh}^{(\mathrm{s})}\right)\tra
    		\end{pmatrix}\tra
    	\label{eq:Vek_Q}
    \fp
    	}{Vek_Fnl}
        \markit{Herein, $(~)\tra$ denotes the transpose.} Substitution of \erefs{Fourier_q}-\erefo{Fourier_ddq} into \eref{EquationOfMotion_gen} generally produces an error term considering that only a limited number of harmonics is taken into account. It is then required that the original equation of motion is weakly enforced with respect to suitable weighting functions. Following the Galerkin idea, the base functions are utilized for this projection, which is known as Fourier-Galerkin projection \zo{urab1965}, and results in the following set of nonlinear algebraic equations,
    	\e{
    		\mm{S}\left(\omega\right)\tilde{\mm{Q}}+\tilde{\mm{F}}_{\mathrm{nl}}\left(\tilde{\mm{Q}}\right)-\tilde{\mm{F}}_{\mathrm{ex}}=:\mm{\tilde R}\left(\tilde{\mm{Q}},\omega\right)\stackrel{!}\to 0\fp
    	}{Vek_R}
        The Fourier coefficients of the forces, Herein, $\tilde{\mm{F}}_{\mathrm{nl}}$ and $\tilde{\mm{F}}_{\mathrm{ex}}$ take a form analogous to \eref{Vek_Q}. In \eref{Vek_R}, $\mm{\tilde R}$ is the residual function, of which a zero characterizes a forced response approximation for a certain frequency $\omega$. $\mm{S}(\omega)$ is the dynamic stiffness matrix and can be expressed as
    	\e{
    		\mm{S}(\omega)=\mm{D}^{2}(\omega)\otimes\mm{M}+\mm{D}^{1}(\omega)\otimes\mm{C}+\mm{D}^{0}(\omega)\otimes\mm{K}\fk
    	}{Mat_S}
    	with the frequency-domain derivative matrix
    	\e{
    		\mm{D}(\omega)=
    		\begin{pmatrix}
    			0 & 0 & 0 & \hdots & 0 & 0 \\
    			0 & 0 & 1\cdot\omega & \hdots & 0 & 0 \\
    			0 & -1\cdot\omega & 0 & \hdots & 0 & 0 \\
    			\vdots & \vdots & \vdots & \ddots & \vdots & \vdots \\
    			0 & 0 & 0 & \hdots & 0 & \nh\cdot\omega \\
    			0 & 0 & 0 & \hdots & -\nh\cdot\omega & 0 \\
    		\end{pmatrix}\fp
    	}{Mat_D}
    	and $\otimes$ is the \name{Kronecker} product.\\
        In \eref{Vek_R}, the coefficients of corresponding sine and cosine functions are thus enforced to add up to zero, that is they are \myquote{balanced}, hence the appropriate name \myquote{harmonic balance method}. While $\tilde{\mm{F}}_{\mathrm{ex}}$ is usually given, the values of $\tilde{\mm{Q}}$ and $\tilde{\mm{F}}_{\mathrm{nl}}$ are unknown and are determined as the solution of \eref{Vek_R}. The solution is usually computed numerically and iteratively. The iterative scheme starts with a well-chosen vector $\tilde{\mm{Q}}_{\left(0\right)}$. A crucial task is the computation of the fourier coefficients $\tilde{\mm{F}}_{\mathrm{nl}}$ of the nonlinear forces. Closed-form expressions of the nonlinear forces in the frequency domain are only available in special cases. In the general case, the nonlinear forces can be more easily evaluated in the time domain. For this purpose, the Alternating Frequency / Time (AFT) method \zo{came1989,cardona1998} is very popular. This method can be summarized as
\e{\tilde{\mm{F}}_{\mathrm{nl}} = \mathrm{DFT}\left[~\fnl\left(\mathrm{iDFT}\left[\tilde{\mm{Q}}\right]\right)~\right]\fk}{aft}
        The estimated coordinates defined by $\tilde{\mm{Q}}$ are transformed to the time domain, using \eref{Fourier_q}, to obtain the temporal evolution $\mm{q}(t)$ and its time derivative. The coordinates are commonly evaluated at discrete points in time only, using the inverse discrete Fourier transform (iDFT). Owing to periodicity, it is sufficient to consider only a single period. The relation $\fnl\left(\mm{q}\left(t\right),\dot{\mm{q}}\left(t\right), \lambda\right)$ can then be evaluated in the time domain in a straight-forward manner. It is then possible to calculate $\tilde{\mm{F}}_{\mathrm{nl}}$ using the discrete fourier transform (DFT). \eref{Vek_R} yields the first residual $\mm{\tilde R}_{\left(0\right)}$. Corrections are then applied to the vector of unknowns to get closer to the solution. This is usually done with gradient-based methods, \eg by taking the Newton step,
    	\e{
    \tilde{\mm{Q}}_{\left(i+1\right)}=\tilde{\mm{Q}}_{\left(i\right)}-\left(\frac{\partial\mm{\tilde R}_{\left(i\right)}}{\partial\tilde{\mm{Q}}_{\left(i\right)}}\right)^{-1}\cdot\mm{\tilde R}_{\left(i\right)}
    	}{Vek_Qi+1}
    	This procedure is continued until $\Vert\mm{\tilde R}_{\left(i\right)}\Vert<\epsilon$ with a small, real-valued tolerance $\epsilon>0$.


    \subsection{The \phaselagmethod\label{sec:phaselag}}
\nc{\lammin}{\lambda_{\mathrm{min}}}
\nc{\lammax}{\lambda_{\mathrm{max}}}
\nc{\omres}{\omega_{\mathrm{res}}}
\nc{\philinres}{\phi_{\mathrm{res,lin}}}
Of particular interest in the design of externally forced systems are resonances. In this study, we limit the discussion to periodic, steady-state vibrations in the presence of harmonic forcing of the frequency $\omega$, such that $\mm{F}_{\mathrm{ex},0}=\mm 0$ and $\mm{F}_{\mathrm{ex},n}^{(\mathrm{c,s})}=\mm 0$ $\forall n>1$. We refer to a resonance as a local maximum of the frequency response curve in the amplitude-frequency plane. It should be noted that in a multi-DOF system, the generalized coordinates do not necessarily reach their maximum amplitudes at the same frequency. Hence, an appropriate, representative coordinate needs to be specified to uniquely define a resonance. Moreover, the notion of amplitude needs to be properly defined in the light of multi-harmonic periodic motions. In this study, we assume that the fundamental harmonic component dominates the forced vibration response. In this \textit{weakly-nonlinear regime}, it is reasonable to define the magnitude of the fundamental harmonic as amplitude,
\e{\hat q_k := \left|Q_1^{(\mathrm{c})}\left[k\right]+\ii Q_1^{(\mathrm{s})}\left[k\right]\right|\fp}{amplitude}
Herein, $\left[k\right]$ denotes the $k$-th component of a vector, which is associated with the coordinate $k$. In general, the maximum or peak-to-peak value within a period of oscillation are interesting alternative definitions. Moreover, the dedicated application range are lightly damped systems with well-spaced modes. The focus on light damping is reasonable since only in this case the condition of resonance is particularly critical. Also, nonlinear systems with closely-spaced modes may exhibit rather unique behavior, and therefore require special treatment.\\
In what follows, a strategy is presented for the direct computation of the resonance curve in the parameter range $\lammin\leq\lambda\leq\lammax$. The key idea is to compute only those solutions of the dynamic equilibrium given in \eref{Vek_R} that satisfy an appropriate resonance condition. Whereas in a conventional forced response analysis, the excitation frequency $\omega$ is specified, this quantity becomes the unknown resonance frequency $\omres$. In fact, the resonance condition defines the relation between the resonance frequency and the resonant vibration response. In the proposed \phaselagmethod, the resonance condition is based on the phase lag between response and excitation.\\
In the neighborhood of a single resonance, the phase lag $\Delta\phi$ between response and excitation undergoes a change of about $180^\circ$, see \fref{af-plf}. In the linear setting and without damping the phase lag at resonance is $-90^\circ$. The phase lag of a linear system can be calculated analytically. In the nonlinear case the definition of phase is not straight forward and the analytical calculation is in general not possible. Moreover, multiple harmonics are generally present in the vibration response. As mentioned above, we are mainly interested in the weakly nonlinear regime around an isolated resonance, where the fundamental harmonic component of the vibration response dominates. It is therefore reasonable to consider only the fundamental harmonic in the definition of the phase lag, and to assume that the phase lag remains close to the one in the linear case. Instead of imposing a condition explicitly on the phase lag between response and excitation, it is sufficient to consider the absolute phase $\phi$ of the response (since the absolute phase of the excitation is invariant by definition). Hence, the resonance condition for the resonance frequency $\omres$ can be expressed as
			\e{
				\phi\left(\lambda,\omres\right)\approx\phi_{\mathrm{res,lin}}\:\:\:\:\:\:\forall\lambda \in \left[\lammin,\lammax\right]\fp
			}{Phi_gleich_phires}
Note that this approximation introduces a certain level of inaccuracy. It should be noted that a similar criterion is widely-used for the investigation of the frequency-energy dependence of nonlinear normal modes \zo{peeters2011}. Of course, this concept is also restricted the weakly nonlinear regime around an isolated resonance of a lightly damped system, as detailed above.\\
Note that the phase lag undergoes a rapid change in the neighborhood of the resonance, especially in the case of lightly damped systems, see \fref{af-plf}. A finite error in the assumed phase in \eref{Phi_gleich_phires} therefore leads to a comparatively small error in the predicted $\omres$. In fact, this is an important reason for the high accuracy that can be achieved with the proposed method, as will be demonstrated later.
\fs[h!]{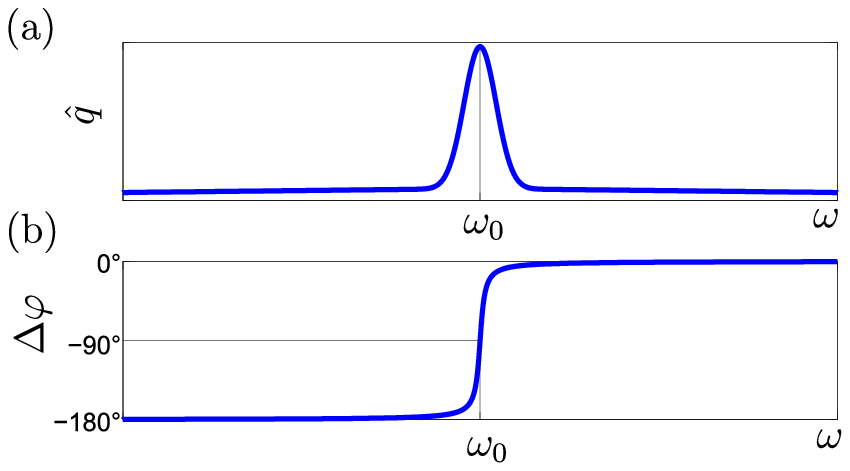}{(a) Amplitude-frequency and (b) phase lag-frequency-curve of a linear system with well-spaced modes}{1.0}
\\
To determine the phase $\phi_{\mathrm{res,lin}}$ in the linear case, \eref{EquationOfMotion_gen} is considered with $\fnl\left(\mm{q}\left(t\right),\lambda\right)=0$. Only the (directly-driven) first harmonic needs to be considered in the linear case. The Fourier coefficients of the generalized coordinates can be expressed in closed form as
\nc{\omreslin}{\omega_{\mathrm{res,lin}}}
\e{\left[ \left(\mm Q_1^{(\mathrm{c})}\right)\tra \left(\mm Q_1^{(\mathrm{s})}\right)\tra \right]\tra = \mm S_1(\omreslin) \left[ \left(\mm F_{\mathrm{ex},1}^{(\mathrm{c})}\right)\tra \left(\mm F_{\mathrm{ex},1}^{(\mathrm{s})}\right)\tra \right]\tra\fp}{phase_lag}
Herein, $\mm S_1$ is the sub-matrix of the dynamic stiffness matrix $\mm S$ associated with the first harmonic,
			\e{
				\mm{S}_1(\omreslin):=
				\begin{pmatrix}
					\mm{K}-\omreslin\cdot\mm{M} && \omreslin\cdot\mm{C} \\
					-\omreslin\cdot\mm{C} && \mm{K}-\omreslin\cdot\mm{M}
				\end{pmatrix}\fk
			}{Pmatrix}
            The dynamic stiffness matrix is evaluated at $\omreslin$ that is the resonance frequency in the linear case. In the presence of light damping, it holds that $\omreslin\approx\omega_0$, and one can instead use the undamped natural frequency $\omega_0$ of a particular mode, which can be calculated even more easily.\\
            In a multi-DOF system, the phase can differ from coordinate to coordinate. If the natural frequencies are well-spaced, and the damping is sufficiently weak, however, the resonant phases of the coordinates remain close to each other. The results should therefore not significantly depend on the choice of the generalized coordinate. The phase of a specified coordinate with index $k$ can be calculated as
			\e{
				\philinres=\textrm{arctan}\left(\frac{Q_1^{\left(s\right)}[k]}{Q_1^{\left(c\right)}[k]}\right)\fp
			}{Phi_res_lin}
The resonance condition can be written in matrix form as
			\e{
				{r}=\mm{g}\tra\mm{X}\fk
			}{r_residuum}
where $\mm X$ is the extended vector of unknowns
\e{
				\mm{X}=
				\begin{pmatrix}
					\tilde{\mm{Q}} \\
					\omres
				\end{pmatrix}\fk
			}{Xtensor}
and $\mm g$ is defined as follows:
			\e{
				\mm{g}\tra =
				\bordermatrix{
					& 1 & \cdots & \ndof+k & \cdots & 2 \cdot \ndof+k & \cdots & \left( 2 \cdot \nh + 1 \right) \cdot \ndof+1 \cr
					& 0 & \cdots & -\tan\left(\philinres\right) & \cdots & 1 & \cdots & 0
				}\fp
			}{Gmatrix}
The extended set of algebraic equations consists of the dynamic equilibrium defined by the residual $\mm{\tilde R}$ in \eref{Vek_R} and the resonance condition $r$ defined in \eref{r_residuum}. An extended residual vector $\mm{R}$ can thus be defined as
			\e{
				\mm{R} =
				\begin{pmatrix}
					\mm{\tilde R} \\
					{r}
				\end{pmatrix}\fp
			}{R_res}
			\eref{R_res} can also be solved by a gradient-based method such as the Newton method, as explained in the previous subsection.\\
In order to increase the computational efficiency of the solution process, it is reasonable to analytically determine the \name{Jacobi} matrix and provide it to the gradient-based solver. The derivatives $\frac{\partial\mm{\tilde R}}{\partial\tilde{\mm{Q}}\tra}$ and $\frac{\partial\mm{\tilde R}}{\partial\omega}$ have already been reported in \zo{petr2006b} and are merely repeated here for clarity,
			\ea{ \frac{\partial\mm{R}}{\partial\tilde{\mm{Q}}\tra}=\frac{\partial}{\partial\tilde{\mm{Q}}\tra}\left(\mm{S}\cdot\tilde{\mm{Q}}+\tilde{\mm{F}}_{\mathrm{nl}}-\tilde{\mm{F}}_{\mathrm{ex}}\right)=\mm{S}+\frac{\partial\tilde{\mm{F}}_{\mathrm{nl}}}{\partial\tilde{\mm{Q}}\tra}
			\label{eq:dR_dQ}\fk\\ \frac{\partial\mm{R}}{\partial\omega}=\frac{\partial}{\partial\omega}\left(\mm{S}\cdot\tilde{\mm{Q}}+\tilde{\mm{F}}_{\mathrm{nl}}-\tilde{\mm{F}}_{\mathrm{ex}}\right)=\frac{\partial\mm{S}}{\partial\omega}\cdot\tilde{\mm{Q}}+\frac{\partial\tilde{\mm{F}}_{\mathrm{nl}}}{\partial\omega}
\fk			
}{dR_dOm}
with
			\ea{
\frac{\partial\mm{S}}{\partial\omega}=\left(2\cdot\mm{D}\cdot\frac{\partial\mm{D}}{\partial\omega}\right)\otimes\mm{M}+\frac{\partial\mm{D}}{\partial\omega}\otimes\mm{C}
			\label{eq:dS_dOm}\fk\\
				\frac{\partial\mm{D}}{\partial\omega}=
				\begin{pmatrix}
					0 & 0 & 0 & \hdots & 0 & 0 \\
					0 & 0 & 1 & \hdots & 0 & 0 \\
					0 & -1 & 0 & \hdots & 0 & 0 \\
					\vdots & \vdots & \vdots & \ddots & \vdots & \vdots \\
					0 & 0 & 0 & \hdots & 0 & \nh \\
					0 & 0 & 0 & \hdots & -\nh & 0				
				\end{pmatrix}\fp
			}{dD_dOm}
In addition, the derivatives of the resonance condition is required, which read as follows:
			\e{
				\frac{\partial{r}}{\partial\tilde{\mm{Q}}}=\mm{g}\tra\fk\quad 
				\frac{\partial{r}}{\partial\omega}=0\fp
			}{dr_dOm}


    \subsection{Qualitative and quantitative comparison between \phaselagmethod and \petrovsmethod\label{sec:qualitative}}
Both the \petrovsmethod \zo{petr2006b} and the \phaselagmethod introduced in this study are based on the harmonic balance method. The only difference is the resonance condition. The \petrovsmethod imposes the condition of a horizontal tangent in the amplitude-frequency plane, while the \phaselagmethod imposes a phase condition. While the \petrovsmethod can be applied ad hoc, the \phaselagmethod requires the preliminary calculation of the reference phase $\philinres$ in the linear case.\\
Owing to the different character of the resonance conditions, different orders of derivatives with respect to the unknowns are involved in either the \phaselagmethod or the \petrovsmethod. It should be noted that the critical part is the derivation of the nonlinear forces $\mm{\tilde F}_{\mathrm{nl}}$ with respect to the generalized coordinates $\mm{\tilde Q}$ (in the frequency domain). While the governing algebraic equations in the \petrovsmethod involve first-order derivatives, no derivatives (\ie only \myquote{zeroth-order derivatives}) are required in the \phaselagmethod. In the context of a gradient-based solution process, therefore second-order derivatives are required in conjunction with the \petrovsmethod, whereas only first-order derivatives are required for the \phaselagmethod. If the derivatives are calculated manually, the \petrovsmethod is thus accompanied by a larger analytical preparation effort. This can be an important aspect in practice, where the nonlinearities may take complicated forms. The main advantage of the \phaselagmethod over the \petrovsmethod is its much lower computational effort. This computational benefit is mainly caused by the different orders of derivatives involved in the problem formulation. \markit{In fact, if the number of coordinates involved in the model is sufficiently large, the \petrovsmethod renders infeasible.}\\
To assess this computational benefit quantitatively, the computational complexity of the mathematical operations was analyzed for both methods. The basis for this comparison is a single evaluation of the residual function $\mm R(\mm X)$. Both methods were implemented in Matlab, and the number of multiplications and additions involved in the individual subroutines were determined as a function of the number $\nh$ of harmonics and $\ndof$ of coordinates. The quantities $Z_{\mathrm a}(\nh,\ndof)$ and $Z_{\mathrm m}(\nh,\ndof)$ represent the number of operations that are required for the \petrovsmethod divided by the according number for the \phaselagmethod, for additions and multiplications, respectively. They take the following form:
\ea{Z_{\mathrm a}(\nh,\ndof) =& \frac{8 N_{h}^{3}+4 N_{h}^{2} \left(5 N_{DOF}^{2}+2\right)}{N_{DOF} \left(2 N_{h}+1\right)-1} + \nonumber\\
&\frac{2 N_{h} \left(10 N_{DOF}^{2}-3 N_{DOF}+1\right)+5 N_{DOF}^{2}-3 N_{DOF}-1}{N_{DOF} \left(2 N_{h}+1\right)-1}\fk\\
Z_{\mathrm m}(\nh,\ndof) = &\frac{\left[2 N_{h}+1\right] \left[N_{DOF}^{3} \left(6 N_{h}+3\right)+2 N_{DOF}^{2}+4 N_{h} \left(N_{h}+1\right)+2\right]}{N_{DOF}}\fp
}{complexity_full}
For $\ndof,\nh\gg 1$ and $\ndof\gg\nh$, the leading orders of complexity can be identified as
\ea{Z_{\mathrm a}(\nh,\ndof):~~ \mathcal O(10\nh\ndof)\fk\\
Z_{\mathrm m}(\nh,\ndof):~~ \mathcal O(12\nh^2\ndof^2)\fp}{complexity}
Hence, the computational benefit of the \phaselagmethod in relation to the \petrovsmethod increases considerably with the harmonic order $\nh$ and the number of generalized coordinates $\ndof$.\\
The \petrovsmethod enforces a horizontal tangent in the amplitude-frequency plane. This is the necessary condition of a local maximum, but also for a local minimum. It is thus possible that the method traces a minimum instead of a maximum, or even a saddle point. This can be avoided by choosing an appropriate initial condition. A problem arises if secondary maxima emerge when the parameter $\lambda$ is varied, as it is shown in \zo{krac2012b}. We refer to a maximum as \textit{secondary maximum} if it is not present in the linear case. Such secondary maxima can be encountered in the strongly nonlinear regime, for instance in the case of nonlinear modal interactions. Owing to the local character of the method, only the initially considered extremum is traced, and the possible occurrence of additional maxima is not detected. Hence, it cannot be ensured that the traced local maximum is actually the global one in the considered frequency range. This is an important conceptual shortcoming of both the \petrovsmethod and the \phaselagmethod. Both methods are therefore limited to frequency and parameter ranges where only an isolated resonance occurs. Of course, this problem can be overcome by computing frequency responses for intermediate values of $\lambda$, identification of all local maxima and their individual tracing.\\
The main shortcoming of the \phaselagmethod compared to the \petrovsmethod is its smaller range of applicability due to the underlying simplifying assumptions. In particular, the method is based on the assumption that the resonant phase lag between response and excitation does not change with the parameter $\lambda$. This is only an approximation, and thus the computed solution points not necessarily coincide with the maxima in the amplitude-frequency plane. Significant errors are expected in the presence of larger damping. In this case, the phase lag is less sensitive with respect to the excitation frequency $\omega$. Hence, even small differences between the assumed and the actual phase at resonance will lead to considerable deviations between predicted and actual resonance frequency $\omres$. In principle, this could be overcome by computing frequency responses for intermediate values of $\lambda$, and updating the value $\philinres$ by its actual value at resonance. Just like the \petrovsmethod, the \phaselagmethod is a local method, and therefore shares the above mentioned limitation regarding the occurrence of secondary maxima. In contrast to the \petrovsmethod, however, the \phaselagmethod can in general not be applied to maxima that are not present in the linear case. This is an important conceptual limitation of the method. Assuming that the phase lag at a secondary maximum does not vary with $\lambda$, the \phaselagmethod could be extended accordingly, by replacing $\philinres$ by the phase identified from an appropriate frequency response computation. This extension is, however, considered as beyond the scope of this work.

\section{Application to a two-DOF oscillator with cubic spring\label{sec:application_twodof}}
	\fw[h!]{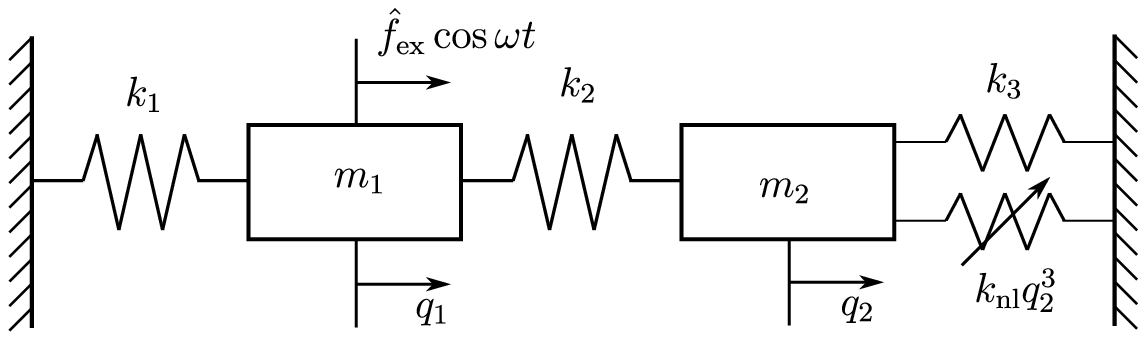}{Nonlinear mechanical system with two DOFs and a cubic spring and external forcing}
In this section, the \phaselagmethod is applied to a numerical example in order to assess its accuracy and computational effort. For reference, the \petrovsmethod is applied as well. Moreover, it is demonstrated how the \phaselagmethod can help to gain insight into the complicated vibration behavior of nonlinear dynamical systems.\\
Consider the nonlinear system of two oscillators and a cubic spring, as depicted in \fref{2dof_qf}. Its equation of motion can be expressed in the form of \eref{EquationOfMotion_gen}, where the vector of coordinates is $\mm q = \left[q_1\,\, q_2\right]\tra$. The mass and stiffness matrices read
\e{
    	\mm{M}=
    	\begin{pmatrix}
    		m_{1} & 0 \\
    		0 & m_{2}
    	\end{pmatrix}\fk\quad 
    	\mm{K}=
    	\begin{pmatrix}
    		k_{1}+k_{2} & -k_{2} \\
    		-k_{2} & k_{2}+k_{3}
    	\end{pmatrix}\fp  
    }{Mat_K}
    The parameters are specified as $m_{1}=m_{2}=1$ and $k_{1}=k_{2}=k_{3}=1$. In addition, a linear, viscous damping is considered, and described in terms of the damping matrix $\mm C$. Proportional damping is assumed with
    \e{
    	\mm{C}=\frac{2D_1}{\omega_{1}}\cdot\mm{K}\fk
    }{Mat_C}
    where $\omega_1$ is the first natural frequency of the linearized system and the modal damping ratio $D_1$ was specified as $D_{1}=1\%$.
    The harmonic excitation is defined as
    \e{
    	\fex=
    	\begin{pmatrix}
    		\hat{f}_{\mathrm{ex}}\cdot\cos\left(\omega t\right) \\
    		0
    	\end{pmatrix}\fk
    }{f_ex}
    with the amplitude $\hat{f}_{ex}=2$.
    The nonlinear spring has a cubic force-deformation behavior and acts on the coordinate $q_{2}$; \ie,
    \e{
    	\fnl=
		\begin{pmatrix}
			0 \\
			k_{\mathrm{nl}}\cdot q_{2}^{3}
		\end{pmatrix}\fp
    }{f_nl}
    The stiffness of the cubic spring $k_{\mathrm{nl}}$ will be varied. Throughout the numerical study, the harmonic order $\nh=3$ was used.

    \subsection{Validation of the \phaselagmethod}

    The first resonance of the system is analyzed. The coordinate $q_2$ was selected for the definition of the amplitude; \ie, $k=2$ in \eref{amplitude}. The parameter $\lambda$ was defined as $\lambda:=k_{\mathrm{nl}}$. It was varied in the range $0\leq\lambda\leq 2.0$. For $\lambda=0$ the system is linear, and thus the calculation of the resonance is trivial. This resonance was specified as initial guess in the subsequent iteration process. The resonance curve was computed with both the \phaselagmethod and the \petrovsmethod. Both methods were implemented in Matlab. For the solution of the nonlinear equations the built-in function \myquote{fsolve} was used, which is based on the Newton method. Starting from the known solution for $\lambda=0$, a sequential continuation was carried out in the interval $\left[0,2.0\right]$. The interval was divided into a number of equidistant steps. In order to achieve numerical convergence of the iterative nonlinear solver, it was found that a smaller step size was required for the \petrovsmethod. Therefore, a number of $2,001$ steps was used for the \phaselagmethod, whereas a number of $200,001$ steps was used in the case of the \petrovsmethod.
\fw[h!]{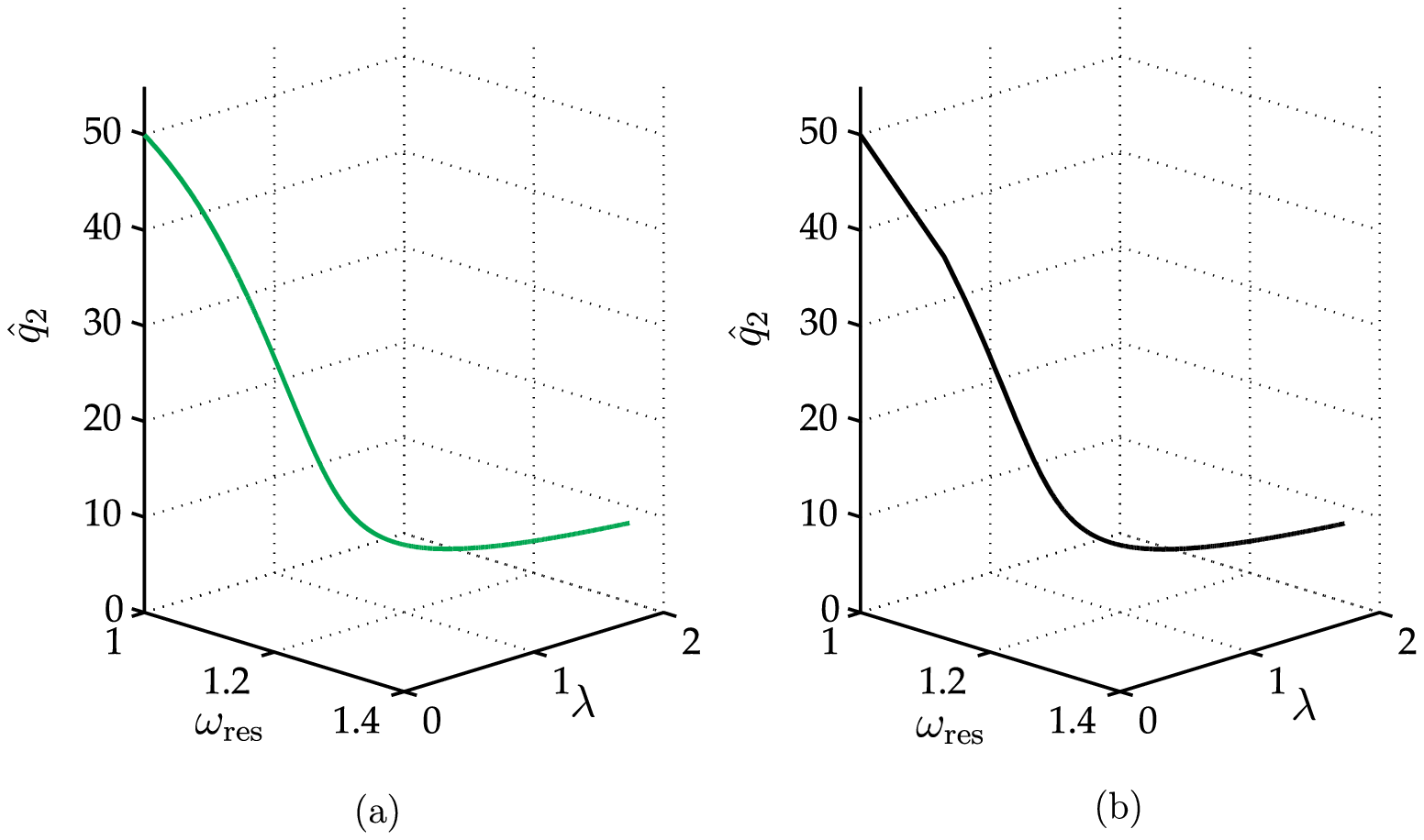}{Relationship between resonance amplitude, frequency and the nonlinear stiffness parameter $\lambda=k_{\mathrm{nl}}$ for the system's first resonance: (a) \petrovsmethod, (b) \phaselagmethod}
\fw[h!]{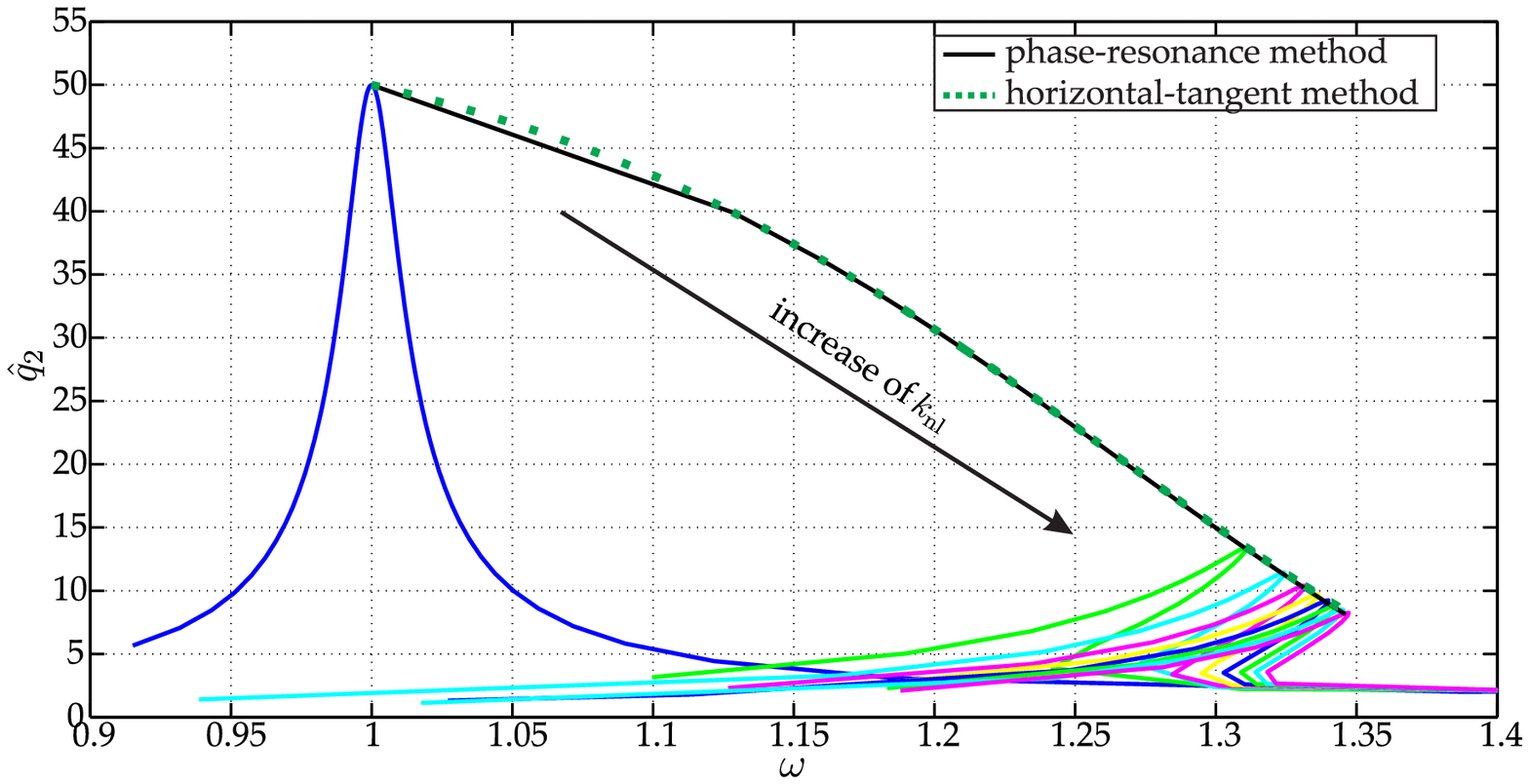}{Resonance curves and frequency responses for several values of $\lambda$}
\fw[h!]{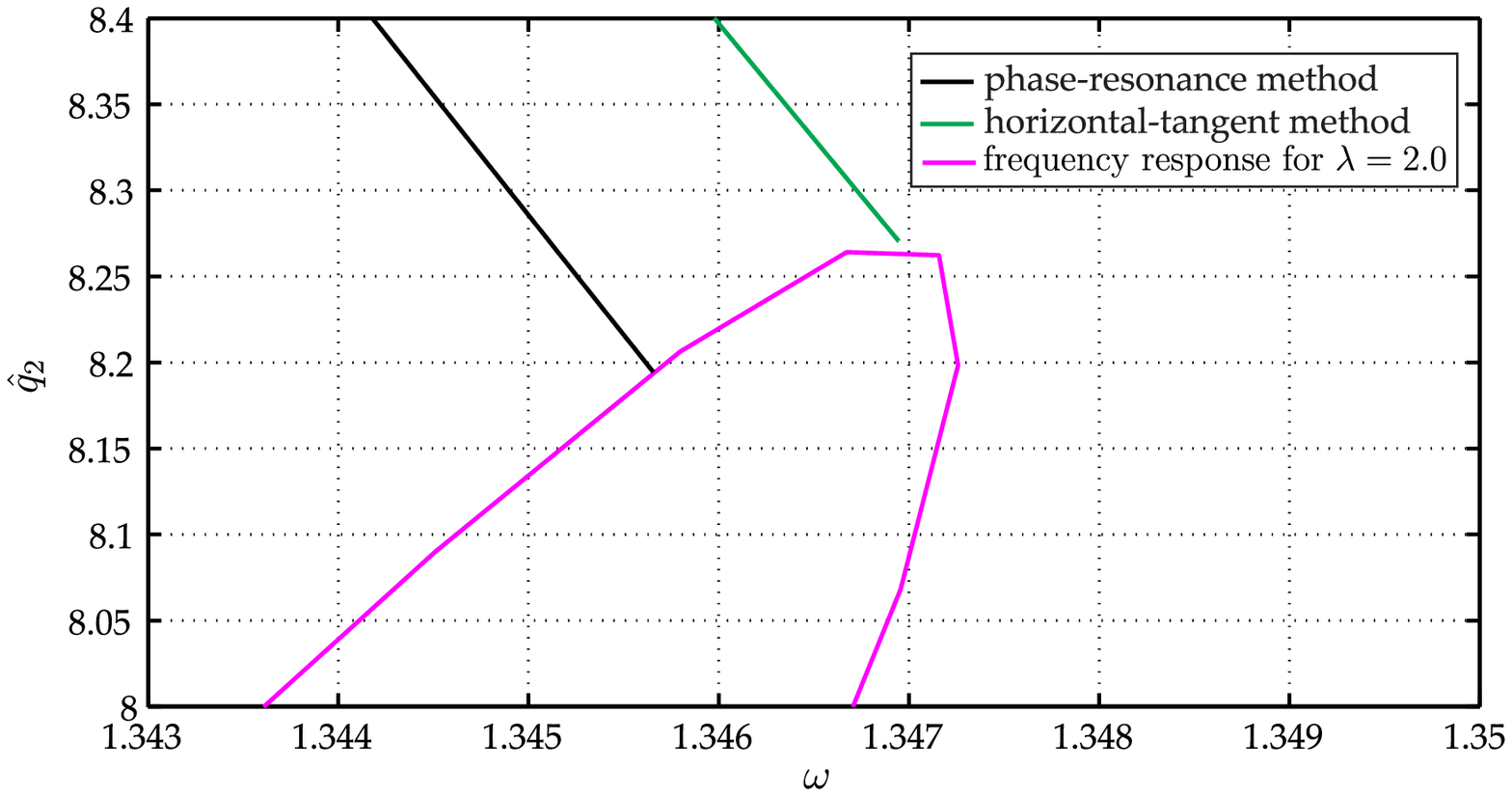}{Zoom-in of \fref{both_methods_and_single} in the neighborhood of the resonance peak for $\lambda=2$}
\\
The relationship between resonance amplitude $\hat q_2(\lambda,\omres)$, resonance frequency $\omres$ and the parameter $\lambda$ is depicted for the \phaselagmethod and the \petrovsmethod in \fref{both_methods_3D}a and b, respectively. The resonance curves are also illustrated in \fref{both_methods_and_single} in the amplitude-frequency plane along with the frequency responses for several values of $\lambda$. It can be stated that results obtained by the \phaselagmethod are in very good agreement with the reference results obtained by the \petrovsmethod. In comparison to different frequency response curves for discrete values of the parameter $\lambda$ both methods provide a good approximation for the resonance curve. The differences seen in \fref{both_methods_3D} and \fref{both_methods_and_single} are mainly due to the different step sizes specified for both methods for the discretization of $\lambda$.
\\
In \fref{both_methods_error_for_l_2}, a zoom of \fref{both_methods_and_single} is depicted in the neighborhood of the resonance peak for $\lambda=2$. This closer look shows that the \petrovsmethod exactly matches the maximum of the frequency response curve. In contrast, the point predicted by the \phaselagmethod is only close to the actual maximum. This confirms that the assumed resonant phase invariance is only an approximation, and thus leads to a certain error. This aspect is investigated further.\\
In \fref{both_methods_phase_lag}, the dependence of the predicted phase lag on $\lambda$ is illustrated. Note that the absolute phase $\phi$ of the response is identical to the phase lag $\Delta\phi$ between response and excitation, $\phi\equiv\Delta\phi$, since the absolute phase of the excitation is zero, see \eref{f_ex}. In the case of the \phaselagmethod, the predicted phase lag is, by definition, a constant. It can be ascertained from the results of the \petrovsmethod, that in fact, the resonant phase changes considerably with the parameter $\lambda$. In spite of this, the presented error of the resonance curve calculated with the \phaselagmethod remains below $1\%$ with regard to the amplitude and below $0.11\%$ with regard to the frequency, in the entire range $0\leq\lambda\leq 2.0$. This can be explained by the low sensitivity of the frequency with respect to the phase near resonance, which leads to a comparatively small error in the predicted $\omres$.
\fw[h!]{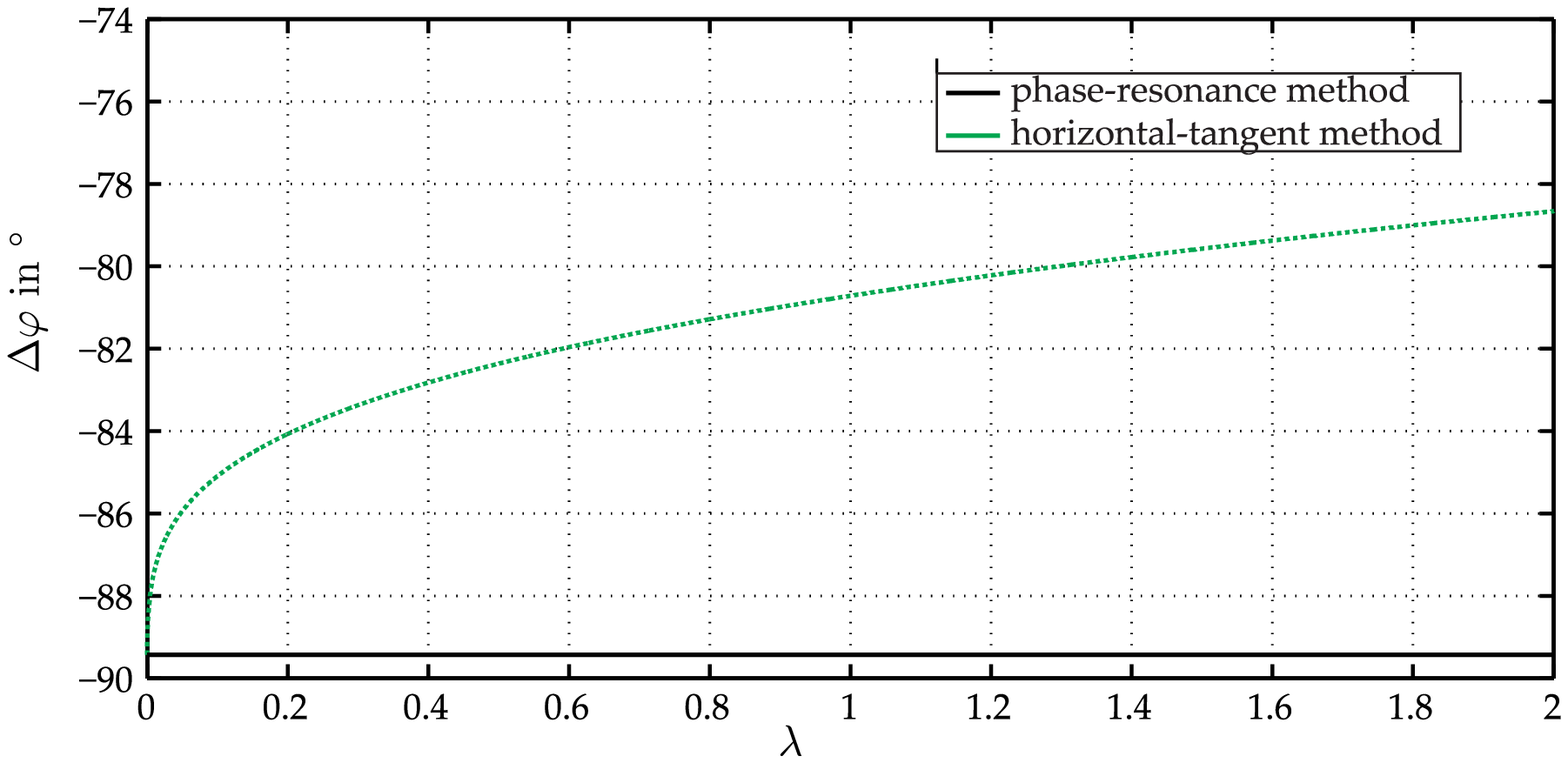}{Dependence of the resonant phase lag on $\lambda$}
    \subsection{Comparison of the computational effort for the numerical example}
%
\tab[tbh]{cccc}{step size & Jacobian & avg. computation time/step & total computation time\\\hline\\ $10^{-5}$ & off & $60.650\mathrm{ms}$ & $606.558694\mathrm{s}$}{Computational effort required in the case of the \petrovsmethod}{zeiten_petrov}
\tab[tbh]{cccc}{step size & Jacobian & avg. computation time/step & total computation time\\\hline\\ $10^{-5}$ & on & $10.980\mathrm{ms}$ & $109.786773\mathrm{s}$ \\ $10^{-5}$ & off & $31.318\mathrm{ms}$ & $313.203478\mathrm{s}$\\ $10^{-3}$ & on & $13.665\mathrm{ms}$ & $1.380140\mathrm{s}$\\ $10^{-3}$ & off & $39.618\mathrm{ms}$ & $4.001173\mathrm{s}$}{Computational effort required in the case of the \phaselagmethod}{zeiten_phaselag}
Next, the computational effort required for the computation of the resonance curve was analyzed for the numerical example. The results are listed in \tref{zeiten_phaselag} and \tref{zeiten_petrov}, for the \phaselagmethod and the \petrovsmethod, respectively. Herein, the average time for the computation of a single point on the resonance curve, as well as the total time required for the entire resonance curve are provided. To this end, the parameter $\lambda$ was varied from $\lambda=0$ to $0.1$. For comparison, the time required for computation of a single frequency response, determined as the average over the ones depicted in \fref{both_methods_and_single}, is $839.639\mathrm{ms}$. Note, however, that this value depends also depends on the step size of the frequency response computation. Details on the computer and software versions are as follows: Operating system Windows 8.1 Pro (64-Bit); Matlab R2013a; CPU Intel(R) Core(TM) i7 860 @ 2.80 GHz, 4 Cores, 8 Threads; RAM 10 GB, (2x4096 MB, 1x2048 MB) @ (Max Bandwidth 667 MHz).\\
As mentioned above, a larger step size was feasible with the \phaselagmethod without any convergence difficulties of the nonlinear solver. Therefore, the computation times were analyzed for two different step sizes in this case: (a) the step size $10^{-5}$ also used in the case of the \petrovsmethod, and (b) a larger step size $10^{-3}$. In the case of the \phaselagmethod, the Jacobian matrix $\partial \mm R/\partial \mm X\tra$ was determined analytically, and also implemented in a variant of the code. In the case of the \petrovsmethod, this is also possible but more complicated, and it was not done in this work. Instead, the Jacobian matrix was approximated using a finite difference scheme, internally within the Matlab function \myquote{fsolve}. To ensure a fair comparison, the analytical Jacobian computation in the case of the \phaselagmethod was switched off, as indicated in \tref{zeiten_phaselag}.\\
By comparing the computation times, one can see that both methods are much faster per step than the computation of a single frequency response curve. If a resolution of $10^{-3}$ is deemed sufficient, it is beneficial to resort to individual frequency response computation rather than using the \petrovsmethod: The computational effort of the \petrovsmethod is seven times larger than for the computation of all $101$ frequency responses, owing to the required smaller step size in the case of the \petrovsmethod. Considering that the computation of individual frequency responses is ad hoc parallelizable, this strategy is clearly preferable over the \petrovsmethod for the considered example.\\
In contrast, the \phaselagmethod is faster than the individual frequency response computation, even for the small step size of $10^{-5}$ and even if the Jacobian is approximated by finite differences. As can be ascertained, however, the use of analytical gradients can greatly accelerate the solution process. The additional analytical preparation effort and the computational overhead for evaluating them thus pays of quickly. More importantly, the better convergence behavior of the \phaselagmethod compared to the \petrovsmethod permits much larger step sizes so that the direct computation of the resonance curve is up to $60$ times fast than the individual computation of frequency responses. However, the above discussed limitations of the \phaselagmethod should be kept in mind.

    \subsection{Investigation of the strongly nonlinear regime}
    If a mechanical system is driven into a highly non-linear regime, it is possible that secondary maxima emerge within a limited frequency range (around a well-spaced mode), which are not present in the linear case. Dynamically, this can be caused \eg by strongly nonlinear modal interactions. Moreover, isolated branches may occur, which are in fact detached from the main solution branch of the frequency response. Hence, for a given $\lambda$, multiple resonances will exist in general. 
    In this subsection, is shown how the \phaselagmethod can be utilized to reveal such isolated branches.\\
Again, the system depicted in \fref{2dof_qf} is used. However, the force $\fex$ now acts on mass $m_{2}$ instead of $m_{1}$. Interestingly, in this case the frequency response exhibits only one maximum for $\lambda < 2.1$, but for $\lambda > 2.1$ at least three maxima are present. An example for this is shown in \fref{lgf_si_fr_re_fu} for $\lambda=5$. This peculiar shape of the amplitude-frequency curve is related to the merging of an isolated branch for a certain value of $\lambda$, in analogy to the results reported \eg in \zo{grolet2014,detroux2014} for similar systems. This behavior is analyzed in the following.
\fw[h!]{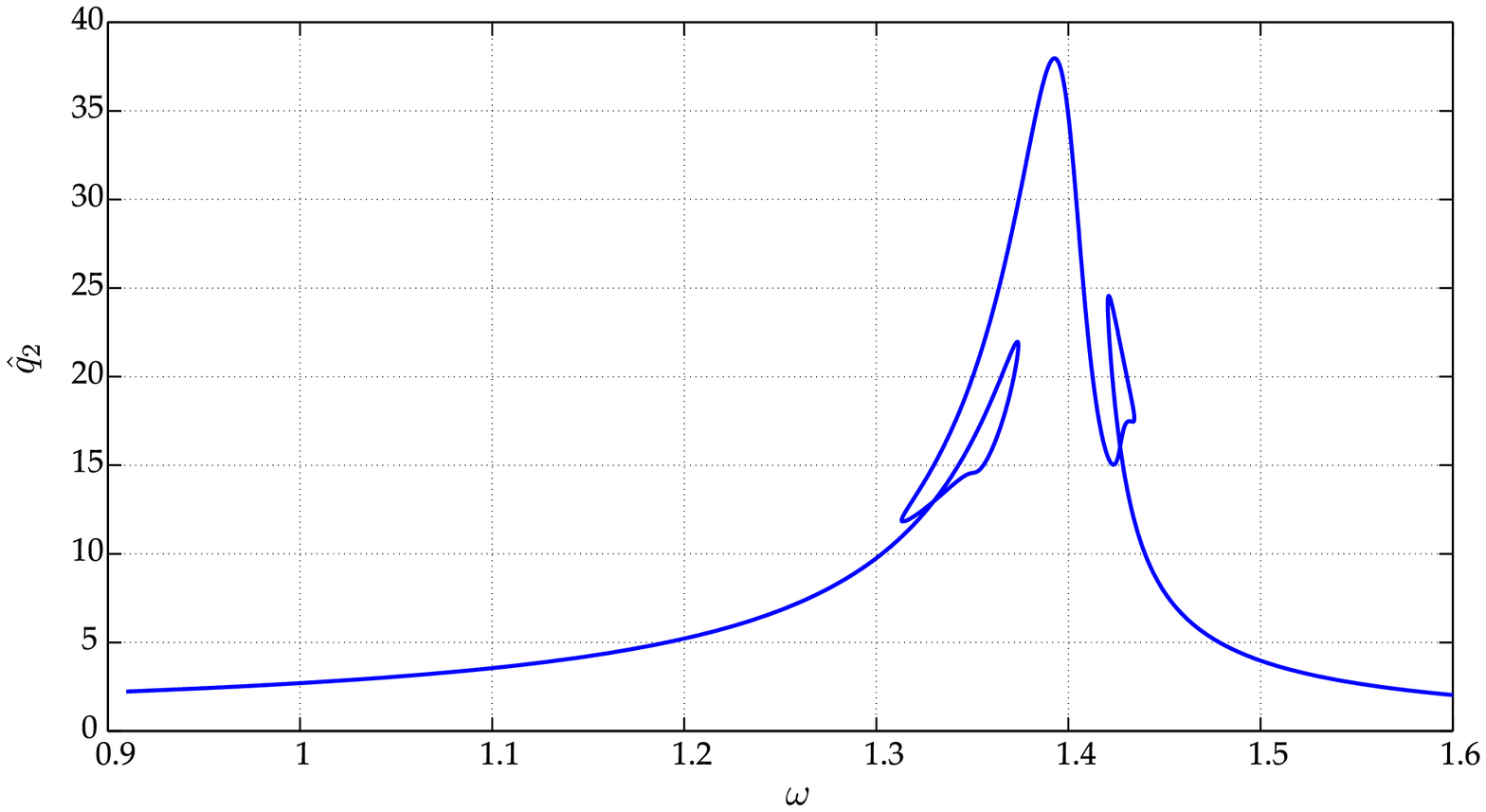}{Frequency response for $\lambda=5$}
\\
During the computation it was noted that the sequential continuation of the solution branch failed, and instead an actual path continuation method had to be employed. The conventional predictor-corrector arc-length continuation method was used in conjunction with an automatic step size adjustment. It was thus possible to compute the resonance curve up to $\lambda=5$ with the \phaselagmethod.
\fw[h!]{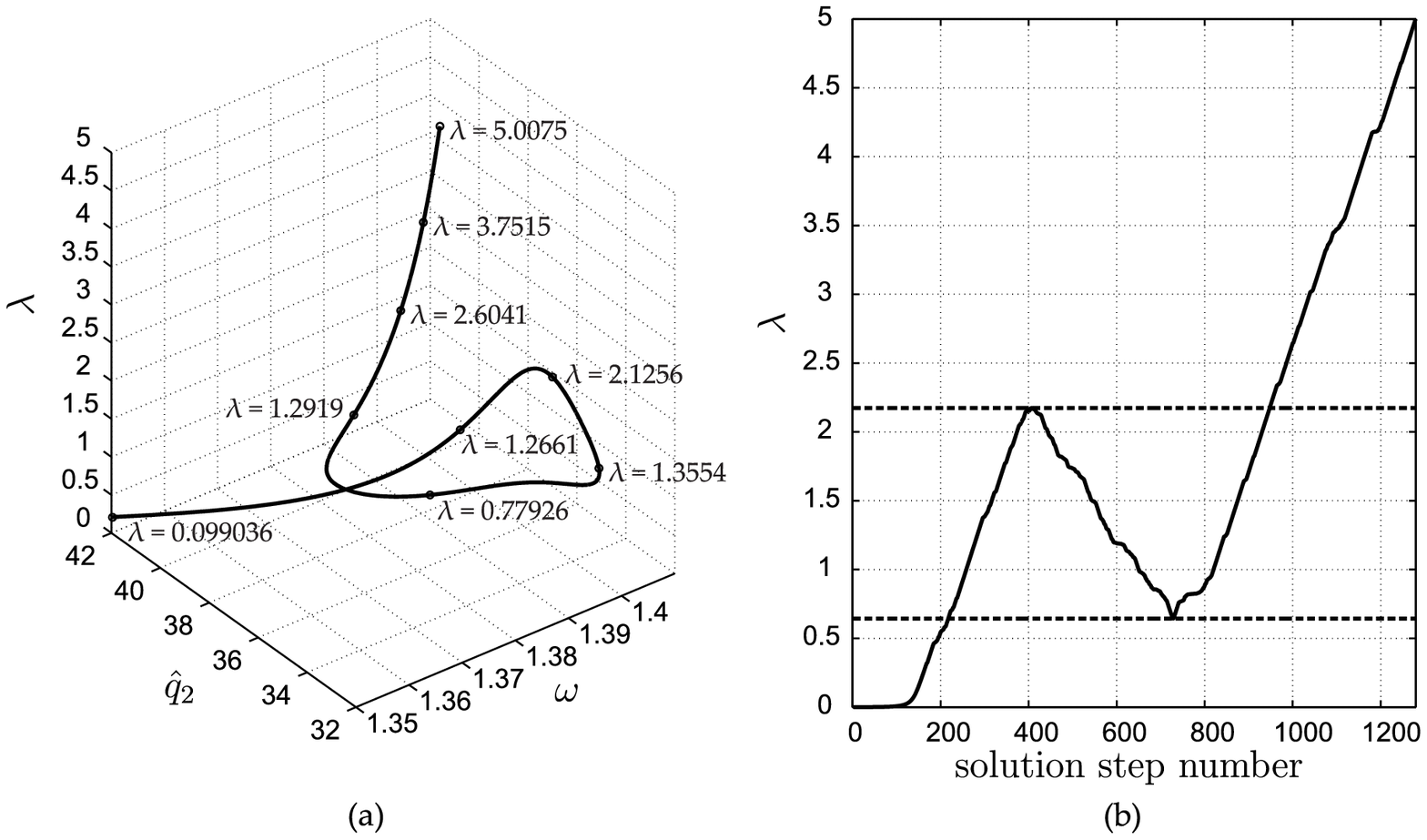}{Resonance curve computed with the \phaselagmethod up to $\lambda=5$: (a) 3D-diagram of the resonance curve as a function of frequency and $\lambda$, (b) shows the changes of the values of $\lambda$ against the number of computation steps}
\fref{lbf_phi_3d} illustrates that $\lambda$ does not increase during the entire continuation, but two turning points are present in the considered range. In the interval $\lambda\in\left[0.64, 2.17\right]$ a second and a third solution for the system of equation are determined. Although the solutions are connected via the (continuous) resonance curve, they are not necessarily located on the same frequency response branch. To demonstrate this, the three solutions for $\lambda=0.93$ were considered, one after the other, and the frequency response was computed starting from either of these points. As it turns out, one of the points is the expected maximum located on the primary branch of the frequency response, as indicated in \fref{isolated_frf}. The other two points are maxima located on an isolated solution branch of the frequency response. This way, the resonance curve computation can be useful to reveal isolated branches of the frequency response.\\
It should be noted that a similar observation was reported by \name{Kuether \etal} in a very recent publication \zo{Kuether.2015}: They investigated a cantilever beam with a cubic spring and also found that isolated branches and main branch of the frequency response are connected via a resonance curve. They determined the resonance curve by means of nonlinear modal analysis of the underlying autonomous and undamped system. The resulting resonance curve thus traces the locus of resonances obtained under variation of the excitation level. It should be emphasized that the approaches discussed in the present work are not limited to the excitation level as a parameter, but any system parameter can be specified (such as the nonlinear stiffness $\lambda:=k_{\mathrm{nl}}$ in our case).
\fw[h!]{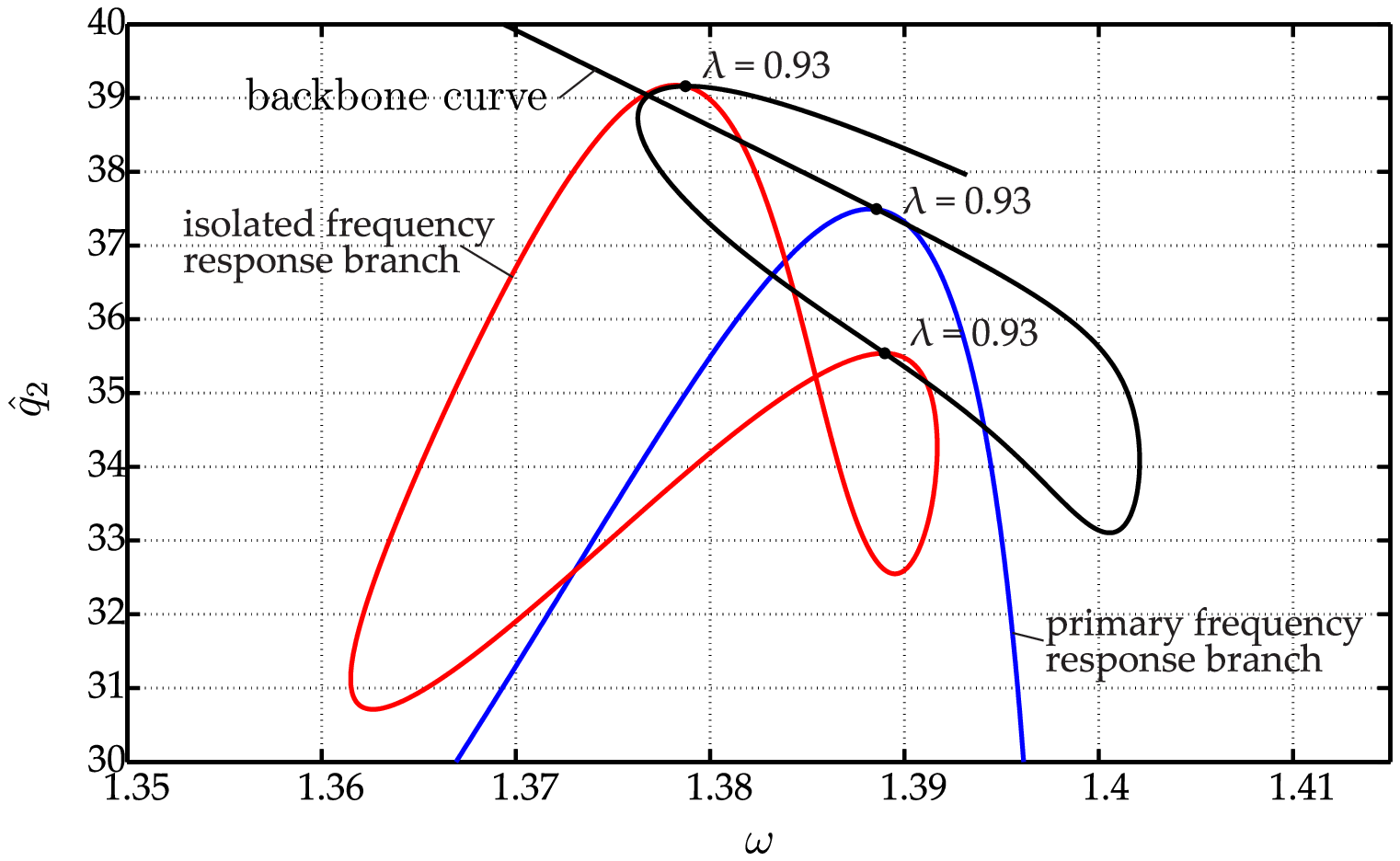}{Isolated and primary branch of the frequency response, connected via the resonance curve (computed by the \phaselagmethod)}
\\
It was also attempted to trace the resonance curve using the \petrovsmethod. However, numerical convergence difficulties were encountered in this case. The numerical convergence could, possibly, be improved by providing an analytically calculated Jacobian. As mentioned earlier, this requires second-order derivatives of the nonlinear forces with respect to the vector of unknowns. This, however, increases the analytical preparation effort by far and was not pursued in this work.\\
From a dynamical point of view, it would be interesting to gain further insight into the behavior in the different regimes. This could be achieved by analyzing the topology and the asymptotic stability of the periodic solutions. Moreover, in addition to the continuation of the resonance curve, an actual bifurcation analysis could be carried out, as \eg proposed in \zo{Cammarano.2014}. Such investigations are, however, are left for future work.

\section{Application to a bladed disk with nonlinear shroud contact\label{sec:application_bladeddisk}}
\fs[h!]{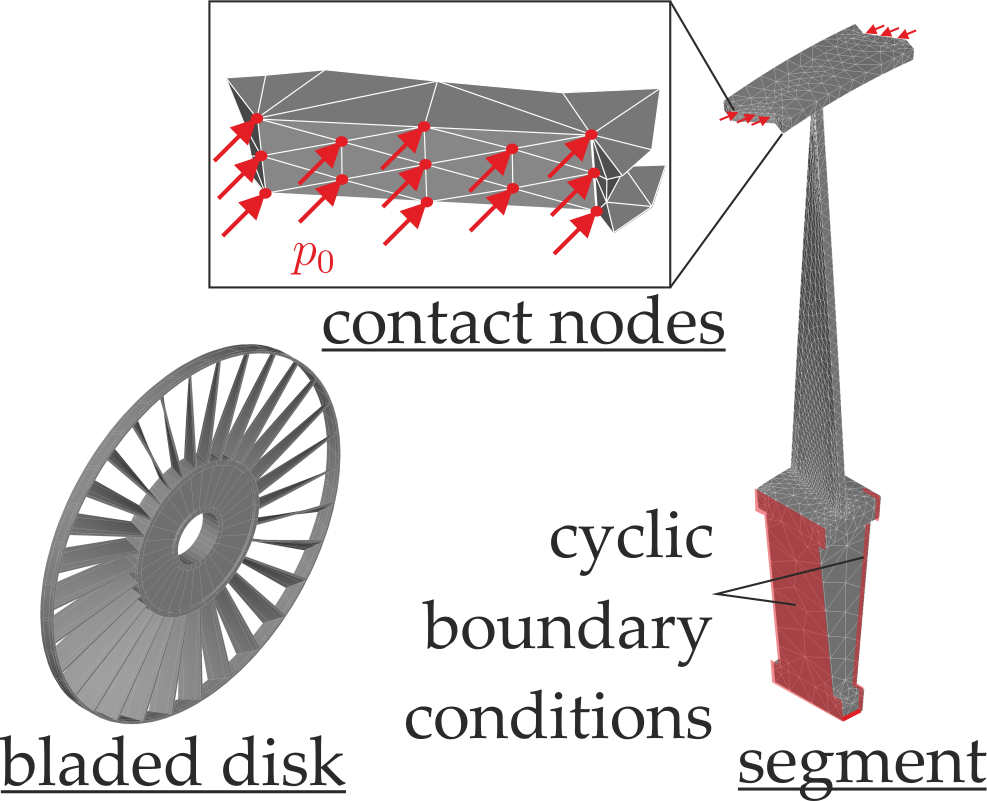}{Finite Element model of a rotationally periodic bladed disk with nonlinear contact at the shroud joints}{1.0}
The application to a two-DOF oscillator in the previous section provided a first validation, insight into the computational effort and the potential usefulness for detecting isolated solution branches. In order to demonstrate that the \phaselagmethod is also useful for more complicated systems, it is applied to a model of a bladed disk with nonlinear shroud contact in this section. Bladed disks are the essential components in turbomachinery applications such as aircraft engines and industrial gas or steam turbines. During operation, bladed disks are exposed to fluctuating fluid forces related to inherent unsteady aerodynamic effects. This type of excitation can cause large vibration levels, and, thus, lead to severe damage. This is particularly true under the condition of resonance. Hence, a major concern in the design of bladed disks is the analysis and mitigation of resonant vibrations. A common strategy to mitigate resonant vibrations is to increase the bladed disks's damping by utilizing the dissipative effects related to dry friction in the mechanical joints. The joints at the tip shrouds, see \fref{bladed_disk} are a common source for friction damping. The accurate vibration prediction in the presence of contact interactions in the mechanical joints is a difficult and computationally demanding endeavor.\\
In the present work, the bladed disk is considered perfectly rotationally periodic, \ie, each of the $30$ segments is assumed to have identical properties. Instead of an actual computation of the aerodynamic loading, the bladed disk is excited by a point force at the center of each blade face, in the axial direction. Along the sectors, the loading is specified as a traveling wave with five waves around the circumference. The traveling wave character of the loading is typical for bladed disks, where the excitation is often caused by the rotation of the bladed disk through a steady, but circumferentially inhomogeneous pressure field. Under the conditions of a rotationally periodic structure and traveling wave type forcing, it is commonly assumed that the steady-state vibrations take the form of traveling waves. This permits one to simplify the problem to only a reference segment with appropriate boundary conditions. In spite of this important reduction, the Finite Element model of a single segment depicted in \fref{bladed_disk} still comprises more than $18,000$ DOFs. For the vibration analysis, component mode synthesis methods are usually employed to considerably reduce the number of generalized coordinates, without significant loss of accuracy. In this work, we use the well-known Craig-Bampton method for this task. To this end, the problem is projected into the subspace spanned by the static constraint modes associated with the relative displacements at the contact nodes in the shroud joint, see \fref{bladed_disk}, and the $50$ leading fixed-interface normal modes. Contact is modeled by (rigid) unilateral contact in the normal direction of the contact interface, and Coulomb's spatial dry friction law in the tangential contact plane. A friction coefficient $0.3$ is specified for both static and dynamic friction. Contact is considered locally at each of the $13$ nodes indicated in \fref{bladed_disk}. A homogeneous initial normal pressure distribution, $p_0$, is considered in the contact interface. The effect of both material and aerodynamical damping is considered as linear modal damping corresponding to a damping ratio of $0.1\%$ for all modes.
\fs[h!]{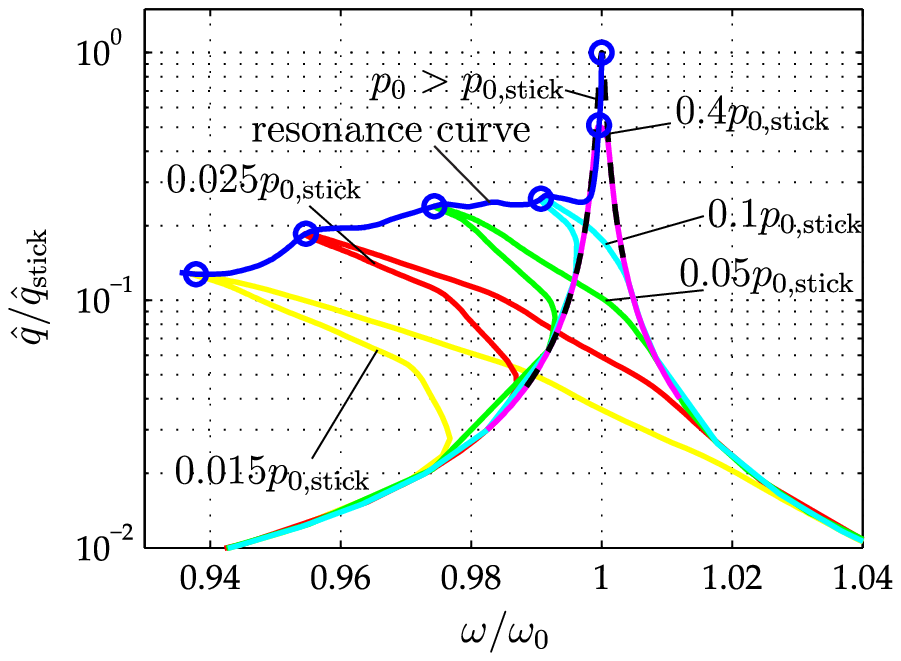}{Resonance curves and frequency responses for several values of the initial normal pressure $p_0$ in the shroud joints}{1.0}
\\
The steady-state, periodic vibrations, are computed using the Harmonic Balance method with a number of $\nh=5$ harmonics. The contact constraints are resolved using the Dynamic Lagrangian method \cite{naci2003}. The results are illustrated in \fref{dpa_ps_compare}, for a variation of the parameter $\lambda=p_0$. For very large initial normal pressures, $p_0$, the contact interface is always sticking. Thus, the system exhibits linear behavior, and reaches comparatively large response levels due to the lack of friction damping. As soon as the preload is sufficiently small $p_0<p_{0,\mathrm{stick}}$, the contact interface locally undergoes periodic stick-slip transitions and liftoff-contact phases. This leads to a considerable softening and damping effect. The strongly nonlinear character of the contact interactions is reflected in the frequency response curves, which are clearly bent towards the left. As can be inferred from \fref{dpa_ps_compare}, the directly computed resonance curve is in very good agreement with the maxima of the individual frequency response curves. The wavelike evolution of the resonance curve is related to the discrete nature of the contact formulation, where the interactions are accounted only at the few nodes indicated in \fref{bladed_disk}. The good agreement between frequency response curves and resonance curve provides a validation of the developed technique for a more sophisticated numerical example.

\section{Conclusions\label{sec:conclusions}}
A novel frequency-domain method was proposed for the direct computation and tracing of resonant vibration responses of harmonically forced nonlinear systems. It is based on the simplified assumption that the phase lag between response and excitation remains invariant under the variation of a parameter. Owing to the typically high sensitivity of the phase with respect to the frequency, this source of inaccuracy often does not result in considerable errors. The numerical results for both the two-DOF oscillator and the bladed disk with shroud contact suggest that the method is useful and accurate in a reasonably wide range. Concept-related limitations are, however, expected for strong damping, closely-spaced modes, and in the case of strongly nonlinear modal interactions, leading to the emergence of secondary resonances not present in the linear case. The latter limitation applies in a similar manner also to the existing alternative, the \petrovsmethod. In general, the \petrovsmethod provides a higher accuracy than the proposed \phaselagmethod. On the other hand, the \phaselagmethod exhibits superior numerical stability, permitted larger step sizes in the continuation and is computationally much less expensive. Being a zeroth-order method, the \phaselagmethod generally requires less analytical preparation, which is an important practical aspect in the case of generic nonlinearities. Even for the comparatively simple numerical example of the two-DOF oscillator, the \petrovsmethod exhibited such a poor numerical performance that the computation of individual frequency responses was more efficient. In contrast, the \phaselagmethod provided a reduction of the computational cost by a factor of $60$ compared to individual frequency response analyses. \markit{This clearly implies that the \phaselagmethod permits the analysis of systems with a much larger number of degrees of freedom, for which the \petrovsmethod would be infeasible. Finally, the numerical study indicated that the \phaselagmethod has a potential application beyond fulfilling its primary purpose of tracing resonances. As was demonstrated, it is capable of revealing isolated solution branches, and can thus be helpful to gain important insight into the dynamics of nonlinear systems.}

\end{document}